\documentstyle[11pt,epsf]{article}

\title{{\sl FITS} Checksum Proposal}

\author{R.L. Seaman,\thanks{Science Data Systems (IRAF) Group,
National Optical Astronomy Observatory, {\it seaman@noao.edu}}
\and W.D. Pence,\thanks{HEASARC, NASA Goddard Space Flight Center,
{\it pence@tetra.gsfc.nasa.gov}}
\and A.H. Rots\thanks{CXC, Harvard-Smithsonian Center for Astrophysics,
{\it arots@head-cfa.harvard.edu}} }

\date{23 May 2002}

\begin{document}

\maketitle

\section{Introduction}

The checksum keywords described here provide an integrity check on the
information contained in {\sl FITS} HDUs.  (Header and Data Units are
the basic components of FITS files, consisting of header keyword
records followed by optional associated data records).  The {\tt
CHECKSUM} keyword is defined to have a value that forces the 32-bit 1's
complement checksum accumulated over all the 2880-byte {\sl FITS}
logical records in the HDU to equal negative 0.  (Note that 1's
complement arithmetic has both positive and negative zero elements).
Verifying that the accumlated checksum is still equal to -0 provides a
fast and fairly reliable way to determine that the HDU has not been
modified by subsequent data processing operations or corrupted while
copying or storing the file on physical media.  The checksum does not
guard against organized transformations or malicious tampering,
however, because simple tranformations, such as rearranging the order
of 32-bit words in the file, do not affect the computed checksum value.
The checksum also does not provide any information on the authenticity
of the file because the {\tt CHECKSUM} keyword can always be updated
after making modifications to the file, leaving no trace that the file
is not the same as the original.  A brief comparison with alternative
checksum algorithms is given in \S A.6.

Two {\sl FITS} keywords are reserved to record the checksum information
in an HDU: {\tt DATASUM} and {\tt CHECKSUM}. Normally both keywords
will be present in the header if either is present, but this is not
required. These keywords apply only to the HDU in which they are
contained.   If the {\tt CHECKSUM} keywords are written in one HDU of a
multi-HDU {\sl FITS} file then it is strongly recommended that they
also be written to every other HDU in the file.  In that case the
checksum accumulated over the entire file will equal -0 as well. It is
recommended that the current date and time be written into the comment
field of both keywords to document when the checksum was computed (or
more precisely, the time that the checksum computation process was
started).

\section{{\tt DATASUM} Keyword}

The value field of the {\tt DATASUM} keyword shall consist of a
character string containing the unsigned integer value of the 32-bit
1's complement checksum of the data records in the HDU (i.e., excluding
the header records).   For this purpose, each 2880-byte {\sl FITS}
logical record should be interpreted as consisting of 720 32-bit
unsigned integers.  The 4 bytes in each integer must be interpreted in
order of decreasing significance where the most significant byte is
first, and the least significant byte is last.  Accumulate the sum of
these integers using 1's complement arithmetic in which any overflow of
the most significant bit is propagated back into the least significant
bit of the sum.

The {\tt DATASUM} value is expressed as a character string (i.e.,
enclosed in single quote characters) because support for the full range
of 32-bit unsigned integer keyword values is problematic in some
software systems.   This string may be padded with non-significant
leading or trailing blank characters or leading zeros.  A string
containing only 1 or more consecutive ASCII blanks may be used to
represent an undefined or unknown value for the {\tt DATASUM} keyword.
The {\tt DATASUM} keyword may be omitted in HDUs that have no data
records, but it is preferable to include the keyword with a value of
0.  Otherwise, a missing {\tt DATASUM} keyword asserts no knowledge of
the checksum of the data records.

\section{{\tt CHECKSUM} Keyword}

The value field of the {\tt CHECKSUM} keyword shall consist of an ASCII
character string whose value forces the 32-bit 1's complement checksum
accumulated over the entire {\sl FITS} HDU to equal negative 0.
There are a vast number of possible character strings that could
satisfy this requirement, but for the sake of consistency and
uniformity it is recommended that the particular 16-character string
generated by the algorithm described in the appendix be used.  A string
containing only 1 or more consecutive ASCII blanks may be used to
represent an undefined or unknown value for the {\tt CHECKSUM}
keyword.
\newpage

\appendix
\section{{\tt CHECKSUM} Keyword Implementation Guidelines}

\subsection{Overview}

Checksums are used to gain confidence in the continued integrity
of all sorts of data.  The normal procedure is to calculate
the checksum of the data on the transmitting side of some communication
channel (including magnetic media) and later to compare that checksum
with the recalculated checksum on the receiving side.  The original
checksum is transmitted separately over the same communication channel.

This scheme works for {\sl FITS} data as for other data, but
separating the checksum from the {\sl FITS} file limits its utility,
especially for archival storage.  It is also hard to see just how to
incorporate a separate checksum into the {\sl FITS} standard.

The internet checksum (ref.\ 5--7) resolves the similar problem of
embedding a checksum within each IP packet by forcing the 1's
complement checksum of the entire packet to equal zero.  This is
accomplished by writing the complement of the calculated checksum into
each packet instead of the checksum itself.

A 1's complement checksum is preferable to a
2's complement checksum (as used by the Unix {\tt sum} command, for
example), since overflow bits are permuted back into the sum and
therefore all bit positions are sampled evenly.  A 32-bit sum is as
quick and easy to calculate as a 16 bit sum due to this symmetry,
providing greater sensitivity to errors (see {\S}A.6).

Arranging to write a binary number into a {\sl FITS} file is unattractive
and limiting.  However, the properties of commutativity and associativity
that make the internet checksum work in the first place, also make it
possible to generalize the technique with an ASCII encoding that may
be embedded within a {\sl FITS} header keyword (ref.\ 1).

Although it is advantageous to store the checksum complement within
each HDU, the only place where that can be done, without violating the
standard, is in the header.  As a consequence, one has to be cognizant of
the fact that this mechanism is not practical for application in
situations where a FITS file is being created dynamically onto a
streaming medium: at the point in time when the header is being
written the value of DATASUM is not yet known, and when DATASUM is
known the header cannot be modified anymore.

\subsection{Recommended {\tt CHECKSUM} Keyword Implementation}

The recommended {\tt CHECKSUM} keyword algorithm described here
generates a 16-character ASCII string that forces the 32-bit 1's
complement checksum accumulated over the entire {\sl FITS} HDU to equal
negative 0 (all 32 bits equal to 1).  In addition, this string will
only contain alphanumeric characters within the ranges \mbox{0--9},
\mbox{A--Z}, and \mbox{a--z} to promote human readability and
transcription.  This {\tt CHECKSUM} keyword value must be expressed in
fixed format, with the starting single quote character in column 11 and
the ending single quote character in column 28 of the {\sl FITS}
keyword record, because the relative placement of the value string
within the keyword record affects the computed HDU checksum.  The steps
in the algorithm are as follows:

\begin{enumerate}
\item
Write the {\tt CHECKSUM} keyword into the HDU header with an initial
value consisting of 16 ASCII zeros ({\tt '0000000000000000'}) where the
first single quote character is in column 11 of the {\sl FITS} keyword
record.   This specific initialization string is required by the
encoding algorithm described in {\S}A.3.  The final comment field of
the keyword, if any, must also be written at this time.  It is
recommended that the current date and time be recorded in the comment
field to document when the checksum was computed.

\item
Accumulate the 32-bit 1's complement checksum over the {\sl
FITS} logical records that make up the HDU header in the same manner as
was done for the data records by interpreting each 2880-byte 
logical record as 720 32-bit unsigned integers.

\item
Calculate the checksum for the entire HDU by adding (using 1's
complement arithmetic) the checksum accumulated over the header records
to the checksum accumulated over the data records (i.e., the previously 
computed {\tt DATASUM} keyword value).

\item
Compute the bit-wise complement of the 32-bit total HDU checksum value by
replacing all 0 bits with 1 and all 1 bits with 0.

\item
Encode the complement of the HDU checksum into a 16-character ASCII
string using the algorithm described in {\S}A.3.

\item
Replace the initial {\tt CHECKSUM} keyword value with this 16-character
encoded string.  The checksum for the entire HDU will now be equal to
negative 0.

\end{enumerate}

\subsection{Recommended ASCII Encoding Algorithm}
 
The algorithm described here is used to generate an ASCII string which,
when substituted for the value of the {\tt CHECKSUM} keyword, will
force the checksum for the entire HDU to equal negative 0.  It is based
on a fundamental property of 1's complement arithmetic that the sum of
an integer and the negation of that integer (i.e, the bitwise
complement formed by replacing all 0 bits with 1s and all 1 bits with
0s) will equal negative 0 (all bits set to 1).   This principle is
applied here by constructing a 16-character string which, when
interpreted as a byte stream of 4 32-bit integers, has a sum that is
equal to the complement of the sum accumulated over the rest of the
HDU.  This algorithm also ensures that the 16 bytes that make up the 4
integers all have values that correspond to ASCII alpha-numeric
characters in the range \mbox{0--9}, \mbox{A--Z}, and \mbox{a--z}.

\begin{enumerate}
\item
Begin with the 1's complement (replace 0s with 1s and 1s with 0s) of
the 32-bit checksum accumulated over all the FITS records in the HDU
after first initializing the {\tt CHECKSUM} keyword with a fixed-format
string consisting of 16 ASCII zeros ({\tt '0000000000000000'}).

\item
Interpret this complemented 32-bit value as a sequence of 4 unsigned
8-bit integers, A, B, C and D, where A is the most significant byte and
D is the least significant.  Generate a sequence of 4 integers, A1, A2,
A3, A4, that are all equal to A divided by 4 (truncated to an integer
if necessary).  If A is not evenly divisible by 4, add the remainder to
A1.  The key property to note here is that the sum of the 4 new
integers is equal to the original byte value (e.g., A = A1 + A2 + A3 +
A4).  Perform a similar operation on B, C, and D, resulting in a total
of 16 integer values, 4 from each of the original bytes, which should
be rearranged in the following order:

{
\begin{quote}
\small
\begin{verbatim}
A1 B1 C1 D1 A2 B2 C2 D2 A3 B3 C3 D3 A4 B4 C4 D4
\end{verbatim}
\end{quote}
}

Each of these integers represents one of the 16 characters in the final
{\tt CHECKSUM} keyword value.  Note that if this byte stream is
interpreted as 4 32-bit integers, the sum of the integers is equal to
the original complemented checksum value.

\item
Add 48 (hex 30), which is the value of an ASCII zero character, to
each of the 16 integers  generated in the previous step.  This 
places the values in the range of ASCII alphanumeric characters '0'
(ASCII zero) to 'r'.  This offset is effectively subtracted back out of
the checksum when the initial {\tt CHECKSUM} keyword value string of 16
ASCII 0s is replaced with the final encoded checksum value.

\item
To improve human readability and transcription of the string, eliminate
any non-alphanumeric characters by considering the bytes a pair at a
time (e.g., A1 + A2, A3 + A4, B1 + B2, etc.) and repeatedly increment
the first byte in the pair by 1 and decrement the 2nd byte by 1 as
necessary until they  both correspond to the ASCII value of the allowed
alphanumeric characters \mbox{0--9}, \mbox{A--Z}, and \mbox{a--z} shown
in Figure 1.  Note that this operation conserves the value of the sum
of the 4 equivalent 32-bit integers, which is required for use in this
checksum application.

\begin{figure}[hbtp]
\centering
\leavevmode
\epsfxsize=5.5in
\epsfbox{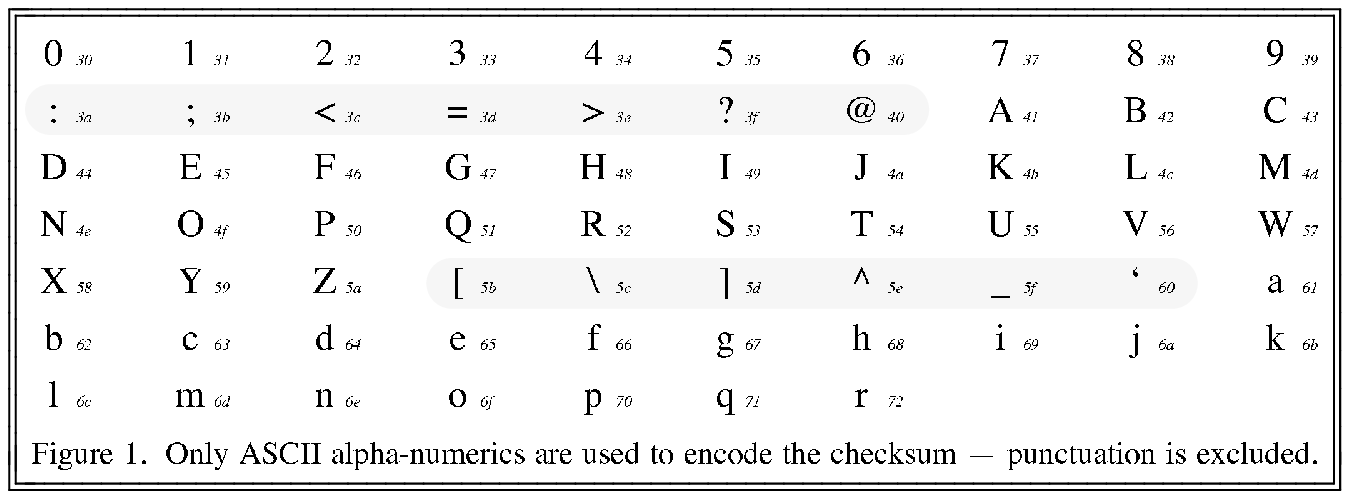}
\end{figure}

\item
Cyclically shift all 16 characters in the string one place to the
right, rotating the last character ({\tt D4}) to the beginning of the
string.  This rotation compensates for the fact that the  fixed format
{\sl FITS} character string values are not aligned on 4-byte word
boundaries in the {\sl FITS} file.  (The first character of the string
starts in column 12 of the header card image, rather than column 13).

\item
Write this string of 16 characters to the value of the {\tt CHECKSUM} keyword,
replacing the initial string of 16 ASCII zeros.

\end{enumerate}

To invert the ASCII encoding, cyclically shift the 16 characters in the
encoded string one place to the left, subtract the hex 30 offset from
each character, and calculate the checksum by interpreting the string
as 4 32-bit unsigned integers.  This can be used, for instance, to read
the value of {\tt CHECKSUM} into the software when verifying or
updating a file.

\subsection{Encoding Example}

This example illustrates the encoding algorithm given in {\S}A.3.
Consider a {\sl FITS} HDU whose 1's complement checksum is 868229149,
which is equivalent to hex {\tt 33C0201D}.  This number was obtained by
accumulating the 32-bit checksum over the header and data
records using 1's complement arithmetic after first initializing the
{\tt CHECKSUM} keyword value to {\tt '0000000000000000'}.   The
complement of the accumulated checksum is 3426738146, which is
equivalent to hex {\tt CC3FDFE2}.  The steps needed to encode this hex
value into ASCII are shown schematically below:

{
\begin{quote}
\small
\begin{verbatim}
    Byte                       Preserve byte alignment
 A  B  C  D      A1 B1 C1 D1  A2 B2 C2 D2  A3 B3 C3 D3  A4 B4 C4 D4

CC 3F DF E2  ->  33 0F 37 38  33 0F 37 38  33 0F 37 38  33 0F 37 38
+ remainder       0  3  3  2

      = hex      33 12 3A 3A  33 0F 37 38  33 0F 37 38  33 0F 37 38
 + 0 offset      30 30 30 30  30 30 30 30  30 30 30 30  30 30 30 30

      = hex      63 42 6A 6A  63 3F 67 68  63 3F 67 68  63 3F 67 68
      ASCII       c  B  j  j   c  ?  g  h   c  ?  g  h   c  ?  g  h

                           Eliminate punctuation characters
 initial values   c  B  j  j   c  ?  g  h   c  ?  g  h   c  ?  g  h 
    .             c  C  j  j   c  >  g  h   c  @  g  h   c  >  g  h
    .             c  D  j  j   c  =  g  h   c  A  g  h   c  =  g  h
    .             c  E  j  j   c  <  g  h   c  B  g  h   c  <  g  h
    .             c  F  j  j   c  ;  g  h   c  C  g  h   c  ;  g  h
    .             c  G  j  j   c  :  g  h   c  D  g  h   c  :  g  h
 final values     c  H  j  j   c  9  g  h   c  E  g  h   c  9  g  h  

 final string     "hcHjjc9ghcEghc9g"   (rotate characters 1 place to the right)
\end{verbatim}
\end{quote}
}

In this example byte B1 (originally ASCII {\tt B}) is shifted higher
(to ASCII {\tt H}) to balance byte B2 (originally ASCII {\tt ?}) being
shifted lower (to ASCII {\tt 9}).  Similarly, bytes B3 and B4 are
shifted by opposing amounts.  This is possible because the two
sequences of ASCII punctuation characters that can occur in encoded
checksums are both preceded and followed by longer sequences of ASCII
alphanumeric characters.   This operation is purely for cosmetic
reasons to improve readability of the final string.

This is how these {\tt CHECKSUM} and {\tt DATASUM} keywords would appear
in a {\sl FITS} header:
{
\begin{quote}
\small
\begin{verbatim}
         1         2         3         4         5         6         7
1234567890123456789012345678901234567890123456789012345678901234567890...

DATASUM = '2503531142'         / Data checksum created 2001-06-28T18:30:45
CHECKSUM= 'hcHjjc9ghcEghc9g'   / HDU checksum created 2001-06-28T18:30:45
\end{verbatim}
\end{quote}
}
\subsection{Incremental Updating of the Checksum}

The symmetry of 1's complement arithmetic also means that after
modifying a {\sl FITS} HDU, the checksum may be incrementally updated
using simple arithmetic without accumulating the checksum for portions
of the file that have not changed.  The new checksum is equal to the
old total checksum plus the checksum accumulated over the modified
records, minus the original checksum for the modified records.

An incremental update provides the mechanism for end-to-end checksum
verification through any number of intermediate processing steps.
By {\sl calculating} rather than {\sl accumulating} the intermediate
checksums, the original checksum test is propagated through to the
final data file.  On the other hand, if a new checksum is accumulated
with each change to the file, no information is preserved about
the file's original state.

The recipe for updating the {\tt CHECKSUM} keyword following some
change to the file is:  \( C' = C - m + m' \), where \( C \) and
\( C' \) represent the file's checksum (that is, the complement
of the {\tt CHECKSUM} keyword) before and after the modification and
\( m \) and \( m' \) are the corresponding checksums for the modified
{\sl FITS} records or keywords only.  Since the {\tt CHECKSUM} keyword
contains the complement of the checksum, the correspondingly complemented
form of the recipe is more directly useful:
\( \tilde{~}C' = \tilde{~}(C + \tilde{~}m + m') \), where \( \tilde{~} \)
(tilde) denotes the (1's) complement operation.  (See ref.\ 5--7.)
Note that the tilde on the right hand side of the equation cannot be
distributed over the contents of the parentheses due to the dual nature
of zero in 1's complement arithmetic (ref.\ 7).

\subsection{Alternate Checksum Algorithms}

There are a variety of checksum schemes (for examples, see ref.\ 4,8--9)
other than the 1's complement algorithm described in this proposal,
although other checksums are significantly more difficult (often
computationally impractical or impossible) to embed in {\sl FITS}
headers in the same fashion.

Checksums, {\sl cyclic redundancy checks} (or {\sl CRCs}, see ref.\ 3
for example), and {\sl message digests} such as MD5 (ref.\ 12) are all
examples of hash functions.  Many possible images will hash to the same
checksum---how many depends on the number of bits in the image versus
the number of bits in the sum.  The utility of a checksum to detect
errors (but not forgeries), to one part in however many bits, depends
on whether it evenly samples the likely errors.

For instance, a 32-bit checksum or CRC each detects the same fraction
of all bit errors (ref.\ 9), missing only \( 1 / 2^{32} \) of all errors
(about 1 out of 4.3 billion) in the limit of long transmissions (the
extra zero of 1's complement arithmetic changes this by
only a small amount).

CRCs and message digests are basically checksums that use higher order
polynomials, thus removing the arithmetic symmetry on which this
proposal relies.  CRCs are tuned to be sensitive to the bursty nature
of communication line noise and will detect all bitstream errors
shorter than the size of the CRC.  Note that the 1's complement sum is
not insensitive to these bit error patterns, it is just not {\it
especially} sensitive to them.  The extra sensitivity of a CRC to burst
errors must come at the expense of lessened sensitivity to other bit
pattern errors (since the total fraction of errors detected remains
the same) and does not necessarily represent the best model for
{\sl FITS} bit errors.  CRCs are also designed to be implemented in
hardware using XOR gates and shift registers that accumulate the function
``on-the-fly'' and emit the CRC {\it after} transmitting the data.
This is not well matched to the {\sl FITS} convention of writing the
metadata as a header which precedes the data records.

\subsubsection{Digital Signatures}

The particular intent of a message digest, on the other hand, is to protect
against human tampering by relying on functions that are computationally
infeasible to spoof.  A message digest should also be much longer than
a simple checksum so that any given message may be assumed to result in
a unique value.

A {\sl digital signature} may be formed by reverse encrypting a message
digest using the private key of a public key encryption pair (ref.\ 13).
A later decryption using the corresponding publically available key
guarantees that the signature could only have been generated by the
holder of the private key, while the message digest uniquely identifies
the document (or image) that was signed.  Support for digital signatures
could be added to the {\sl FITS} standard by defining a {\sl FITS}
extension format to contain the digital signature certificates, or
perhaps by simply embedding them in an appended {\sl FITS} table extension.

There is a tradeoff between the error detection capability of these
algorithms and their speed.  The overhead of a digital signature (or
a software emulated CRC) is larger than a simple checksum,
but may be essential for certain purposes (for instance, archival
storage) in the future.  The checksum defined by this proposal provides
a way to verify {\sl FITS} data against likely random errors, while on
the other hand a full digital signature may be required to protect the
same data against systematic errors, especially human tampering.

\subsubsection{Fletcher's Checksum}

One other checksum algorithm shows some promise of being embeddable in
an ASCII {\sl FITS} header.  This is {\it Fletcher's checksum} (ref.\ 9--11)
which is a variant of the 1's complement checksum that is tuned to trap
bit error patterns similar to those trapped by a CRC.  It is somewhat
slower than the 1's complement checksum and more finicky to implement.
The checksum is divided into two (16 bit) pieces---a straight 1's complement
sum and a higher order sum of the running sums.  The procedure for updating
the two checksum fields (zeroing the checksum of the file) involves solving
a pair of simultaneous equations.  ASCII encoding the checksum would require
an iterative solution spread over the four separate ASCII encoded integer
words (and including the constraint of the hex 30 offset).  Incremental
updating of the checksum would incur a similar penalty for each word of
the {\sl FITS} file that was modified.

The added complexity and overhead of handling Fletcher's checksum
(see ref.\ 10--11) are unwarranted for {\sl FITS}, at least as the default
algorithm, but this checksum is an interesting possibility for binary
applications.  Other checksums are also options in the binary case,
especially if the checksum fields can be located at the end of the file,
which simplifies the arithmetic significantly.

\subsubsection{Error Correcting Algorithms}

Error {\sl correcting} (see ref.\ 2), as opposed to error {\sl detecting},
algorithms are beyond the scope of this proposal, as are non-systematic
codes for either error detection or correction.  {\it Systematic} codes
are those, such as the 1's complement checksum, that require no change
to the data when applied to a message.  Simply appending a checksum to
a file is systematic, as is appending parity or other check bits to each
byte or record of the data without otherwise modifying the data bits.
Codes that are not systematic involve recoding the individual data bits
in some fashion (see the discussion of {\it product} codes in ref.\ 4,
for example).

\newpage

\subsection{Example {\sl C} Code}

\subsubsection{Accumulating the Checksum}

The 1's complement checksum is simple and fast to compute.  This
routine assumes that the input records are a multiple of 4 bytes long
(as is the case for {\sl FITS logical records}), but it is not
difficult to allow for odd length records if necessary.  To use this
routine, first initialize the {\tt CHECKSUM} keyword to {\tt
'0000000000000000'} and initialize {\tt sum32 = 0}, then step
through all the {\sl FITS} logical records in the FITS HDU.

\begin{quote}
\scriptsize
\begin{verbatim}
void checksum (
    unsigned char *buf,  /* Input array of bytes to be checksummed */
                         /*  (interpret as 4-byte unsigned ints)   */
    int length,          /* Length of buf array, in bytes          */
                         /*  (must be multiple of 4)               */
    unsigned int *sum32) /* 32-bit checksum                        */
{
/*
    Increment the input value of sum32 with the 1's complement sum 
    accumulated over the input buf array.        
*/
    unsigned int hi, lo, hicarry, locarry, i;

    /*  Accumulate the sum of the high-order 16 bits and the */
    /*  low-order 16 bits of each 32-bit word, separately.  */
    /*  The first byte in each pair is the most significant. */
    /*  This algorithm works on both big and little endian machines. */

    hi = (*sum32 >> 16);
    lo = *sum32 & 0xFFFF;
    for (i=0; i < length; i+=4) {
        hi += ((buf[i]   << 8) + buf[i+1]);  
        lo += ((buf[i+2] << 8) + buf[i+3]);
    }

    /* fold carry bits from each 16 bit sum into the other sum */
    hicarry = hi >> 16; 
    locarry = lo >> 16;
    while (hicarry || locarry) {
        hi = (hi & 0xFFFF) + locarry;
        lo = (lo & 0xFFFF) + hicarry;
        hicarry = hi >> 16;
        locarry = lo >> 16;
    }

    /* concatenate the full 32-bit value from the 2 halves */
    *sum32 = (hi << 16) + lo;
}
\end{verbatim}
\normalsize
\end{quote}

\newpage
\subsubsection{ASCII Encoding}

This routine encodes the complement of the 32-bit HDU checksum value
into a 16-character string. The byte alignment of the string is
permuted one place to the right for {\sl FITS} to left justify the string
value starting in column 12.

\begin{quote}
\scriptsize
\begin{verbatim}
unsigned int exclude[13] = { 0x3a, 0x3b, 0x3c, 0x3d, 0x3e, 0x3f, 0x40,
                             0x5b, 0x5c, 0x5d, 0x5e, 0x5f, 0x60 };

int offset = 0x30;                          /* ASCII 0 (zero) */
unsigned long mask[4] = { 0xff000000, 0xff0000, 0xff00, 0xff  };

void char_encode (
     unsigned int value,  /* 1's complement of the checksum value */
                          /*     to be encoded                    */
    char *ascii)          /* Output 16-character encoded string   */
{
    int byte, quotient, remainder, ch[4], check, i, j, k;
    char asc[32];

    for (i=0; i < 4; i++) {
        /* each byte becomes four */
        byte = (value & mask[i]) >> ((3 - i) * 8);
        quotient = byte / 4 + offset;
        remainder = byte % 4;
        for (j=0; j < 4; j++)
            ch[j] = quotient;

        ch[0] += remainder;

        for (check=1; check;)           /* avoid ASCII punctuation */
            for (check=0, k=0; k < 13; k++)
                for (j=0; j < 4; j+=2)
                    if (ch[j]==exclude[k] || ch[j+1]==exclude[k]) {
                        ch[j]++;
                        ch[j+1]--;
                        check++;
                    }

        for (j=0; j < 4; j++)           /* assign the bytes */
            asc[4*j+i] = ch[j];
    }

    for (i=0; i < 16; i++)              /* permute the bytes for FITS */
        ascii[i] = asc[(i+15)%16];

    ascii[16] = 0;                      /* terminate the string */
}
\end{verbatim}
\normalsize
\end{quote}

\newpage

\subsection{Acknowledgements}

The authors gratefully acknowledge the many helpful comments from
Barry Schlesinger.

\subsection{References}

\begin{enumerate}
 \item	Seaman, R.L. 1994,
	``{\sl FITS} Checksum Verification in the NOAO Archive'', presented
	at the conference {\sl Astronomical Data Analysis Software and
	Systems IV}, to appear in the {\sl A.S.P. Conf. Ser.}.
 \item	Peterson, W.W. and Weldon Jr., E.J. 1972,
	{\sl Error-Correcting Codes}, Second Edition (MIT Press).
 \item	McNamara, J.E. 1982, {\sl Technical Aspects of Data Communication},
	Second Edition (Digital Press).
 \item	Plummer, W.W. 1978, ``TCP Checksum Function Design'',
	{\sl ACM Computer Communication Review}, {\bf 19}, no. 2, 95-101,
	this is an appendix to {\sl Internet RFC 1071}.
 \item	Braden, R. T., Borman, D.A., and Partridge, C. 1988 (September),
	``Computing the Internet Checksum'',
	{\sl ACM Computer Communication Review}, {\bf 19}, no. 2, 86-94,
	this is {\sl Internet RFC 1071}.
 \item	Mallory, T. and Kullberg, A. 1990 (January),
	``Incremental Updating of the Internet Checksum'',
	{\sl Internet RFC 1141}.
 \item	Rijsinghani, A. (ed.) 1994 (May),
	``Computation of the Internet Checksum via Incremental Update'',
	{\sl Internet RFC 1624}.
 \item	Zweig, J. and Partridge, C. 1990 (March),
	``TCP Alternate Checksum Options'', {\sl Internet RFC 1146}.
 \item	Fletcher, J.G. 1982, ``An Arithmetic Checksum for Serial Transmission'',
	{\sl IEEE Transactions on Communications}, {\bf COM-30}, no. 1, 247-252.
 \item	Nakassis, A. 1988, ``Fletcher's Error Detection Algorithm: How to
	implement it efficiently and how to avoid the most common pitfalls'',
	{\sl ACM Computer Communication Review}, {\bf 18}, no. 5, 63-88.
 \item	Sklower, K. 1989, ``Improving the Efficiency of the OSI Checksum
	Calculation'',
	{\sl ACM Computer Communication Review}, {\bf 19}, no. 5, 32-43.
 \item	Rivest, R. 1992 (April), ``The MD5 Message Digest Algorithm'',
	{\sl Internet RFC 1321}, see also {\sl RFC 1319} and {\sl RFC 1320}.
 \item	Zimmermann, P. 1995, {\sl The Official PGP User's Guide} (MIT Press),
	PGP is available from {\sl http://net-dist.mit.edu/pgp.html} or
	{\sl ftp://ftp.csn.net/mpj/README}, which also provide
	United States export and licensing requirements.
\end{enumerate}

Internet {\it Requests for Comments}, or {\sl RFCs}, are the written
design documents for internet protocols.  They are available at many
locations on the internet, including
{\sl http://www.cis.ohio-state.edu/htbin/rfc/rfc-index.html}.

\end{document}